\documentclass[twocolumn,prb,aps,showpacs,floatfix]{revtex4}

\usepackage[dvips]{graphicx}

\begin{document}
%\draft
%\final

\title{Evidence for short range orbital order in paramagnetic
       insulating (Al,V)$_2$O$_3$}

\author{P. Pfalzer}
\author{J. Will}
\author{A. Nateprov}
\author{M. Klemm}
\author{V. Eyert}
\author{S. Horn}
\affiliation{Institut f\"{u}r Physik, Universtit\"{a}t Augsburg,
Universit\"{a}tsstra{\ss}e 1, 86159 Augsburg, Germany }
\author{A. I. Frenkel}
\affiliation{Physics Department, Yeshiva University, 245 Lexington Avenue, New
York, New York 10016}
\author{S. Calvin}
\affiliation{Naval Research Lab, Code 6340, Washington, DC 20375}
\author{M. L. denBoer}
\affiliation{Department of Physics, Hunter College, City University of New
York, 695 Park Avenue, New York, New York 10021}
\date{\today}

\begin{abstract}
The local structure of (Al$_{0.06}$V$_{0.94}$)$_2$O$_3$ in the paramagnetic
insulating (PI) and antiferromagnetically ordered insulating (AFI) phase has
been investigated using hard and soft x-ray absorption techniques. It is shown
that: 1) on a local scale, the symmetry of the vanadium sites in both the PI
and the AFI phase is the same; and 2) the vanadium $3d$ - oxygen $2p$
hybridization, as gauged by the oxygen $1s$ absorption edge, is the same for
both phases, but distinctly different from the paramagnetic metallic phase of
pure V$_2$O$_3$. These findings can be understood in the context of a recently
proposed model which relates the long range monoclinic distortion of the
antiferromagnetically ordered state to orbital ordering, if orbital short
range order in the PI phase is assumed. The measured anisotropy of the x-ray
absorption spectra is discussed in relation to spin-polarized density
functional calculations.
\end{abstract}


\maketitle

\section*{Introduction}
The metal-insulator transition (MIT) in V$_2$O$_3$ has been intensively
investigated and discussed for many years as an example of a classical Mott
transition. However, such a picture is blurred by the complexity of the
V$_2$O$_3$ phase diagram \cite{McWhan73}\ which involves magnetic and
structural transitions coinciding with the MIT, as a function of temperature,
doping (Cr, Al, Ti), pressure, and oxygen stoichiometry. At room temperature
pure V$_2$O$_3$ is in a paramagnetic metallic (PM) phase which x-ray
diffraction (XRD) shows to be trigonal. At about 180 K there occurs a
transition to an antiferromagnetically insulating (AFI) monoclinic phase.
Doping with Cr or Al results in the formation of a paramagnetic insulating
(PI) phase. The lattice parameters change but long-range trigonal symmetry is
preserved, as indicated by XRD.\cite{McWhan69_1, Spalek90} The role of
electronic correlations in the interplay between changes in the physical
structure, the magnetic properties, and the electronic structure at the
various transitions needs further investigation. It is still not certain
whether the electronic transition is driven by structural changes or
vice-versa. Although recent LDA + DMFT calculations\cite{Held01} show the
importance of electronic correlations in a description of the electronic
structure, a description of the MIT in V$_2$O$_3$ must take into account the
relationship between physical structure and the electronic and magnetic
properties of the system.

The Mott-Hubbard picture of  V$_2$O$_3$, in particular a description in terms
of a one band Hubbard model,\cite{Rozenberg95} is based on a level scheme of
Castellani \textit{et al.}\ for the electronic structure of
V$_2$O$_3$.\cite{Cast78} The crystal field generated by the O octahedron
splits the V $3d$ states into upper e$_g$ and lower $t_{2g}$ states, and the
latter are further split into $a_{1g}$ and $e_g^\pi$ states due to the
trigonal symmetry of the lattice. Interactions between nearest vanadium
neighbors along the $c$-direction (vertical pairs) splits the $a_{1g}$ states
into bonding and antibonding molecular orbitals. According to Castellani
\textit{et al.}\ the bonding $a_{1g}$ orbital is fully occupied while the
antibonding $a_{1g}$ orbital shifts energetically above the $e_g^\pi$ states.
This leaves only one electron per vanadium atom to occupy the $e_g^\pi$
states, leading to orbital degeneracy, which is susceptible to a
degeneracy-lifting process such as the Jahn-Teller effect or orbital ordering.
Indeed, Bao \textit{et al.}\cite{Bao98,Bao97}\ suggested that such orbital
ordering occurs in V$_2$O$_3$. This conclusion was based on neutron scattering
experiments which show disagreement between the propagation vector
characterizing the AFI phase and the propagation vector expected from magnetic
short range order in the PM and PI phases of pure V$_2$O$_3$ and its alloys.
The latter propagation vector is identical to that of a spin density wave in
vanadium-deficient V$_2$O$_3$. The suggested orbital ordering would
distinguish the AFI phase of V$_2$O$_3$ from all the other phases and prevent
a unified description of the MIT in this compound. The validity of
Castellani's model has been called into question by soft x-ray
absorption\cite{Mueller97,Park00}and band structure
calculations.\cite{Ezhov99}Near-edge x-ray absorption fine structure (NEXAFS)
measurements of the V $2p$ and O $1s$ edges provides information on the
unoccupied states near the Fermi level. M\"{u}ller \textit{et al.}\ concluded from
such a NEXAFS study of the different phases of V$_2$O$_3$ and (V,Cr)$_2$O$_3$
that all the insulating phases have, within experimental error, identical
local electronic structures.\cite{Mueller97} In addition, angular resolved
NEXAFS measurements in the metallic and insulating phase of V$_2$O$_3$ are
inconsistent with the assumption that the first excited states are purely
$e_g^\pi$. Rather, the isotropy of the absorption spectra observed in the
metallic phase suggests the first excited states are a mixture of $e_g^\pi$
and $a_{1g}$, while the anisotropy observed in the insulating phases suggests
these states have increased $a_{1g}$ character. These conclusions were
confirmed by LDA+U band structure calculations by Ezhov \textit{et
al.}\cite{Ezhov99}\ and NEXAFS studies by Park \textit{et al.}\cite{Park00}
Using a model fit to the V $2p$, Ezhov \textit{et al.}\ showed that orbital
occupancy of the $e_g^\pi$ states is larger than one and changes at both
MIT's. This renders the one band Hubbard model\cite{Rozenberg95} often applied
to the MIT in V$_2$O$_3$ inadequate.

Important for a unified view of the MIT in V$_2$O$_3$ is the understanding of
the relationship between electronic, magnetic, and structural changes at the
MIT. Recently, a model was proposed\cite{Shiina01,Mila00} that takes into
account degrees of freedom of molecular orbitals formed by vertical V-V pairs
and their interaction within the $ab$-plane, offering a consistent picture of
the magnetic and structural properties of the AFI phase.

In this paper we show using EXAFS and NEXAFS techniques that, on a local
scale, the structural and electronic properties of the AFI and PI phase are
the same. This fact is attributed to short range orbital order in the PI
phase, consistent with recent model calculations\cite{Shiina01, Mila00} and
the characteristics of magnetic short range order observed by neutron
scattering.\cite{Bao98} Spin polarized density functional theory calculations
presented here reflect the anisotropy of the O $2p$ density of states observed
in the insulating phases of the V$_2$O$_3$ phase diagram.

\section*{Experimental and Theoretical Methodology}
Measurements were performed on an Al-doped V$_2$O$_3$ single crystal about 2.5
mm by 1.5 mm by 1 mm grown by chemical transport using TeCl$_4$ as transport
agent and a mixture of V$_2$O$_3$ and Al$_2$O$_3$ powder corresponding to a
nominal concentration x = 0.1 in (V$_{1-x}$Al$_x$)$_2$O$_3$. Energy dispersive
x-ray scattering showed the actual Al concentration in the single crystal was
6 atomic percent, presumably due to lower transport rate of Al compared to
vanadium during crystal growth. XRD showed the expected trigonal structure
with lattice parameters $c_{hex} = 13.81(1)$ \AA\ and  $a_{hex} = 4.985(5)$
\AA, both smaller than those found by Joshi \textit{et al.}\ ($c_{hex} \approx
13.89$ \AA\ and $a_{hex} \approx 5.00$ \AA\ at 6.2 at\%).\cite{Joshi77} The
resulting $c/a$ ratio of 2.7703 is much smaller than in the metallic phase
(2.828). Resistivity measurements, the small value of the $c/a$ ratio, and the
Al concentration show the sample was in the PI phase at room temperature. The
transition from PI to AFI occurred at 165 K as determined by magnetic
susceptibiblity measurements using a Quantum Devices SQUID magnetometer. Laue
diffraction measurements were performed on the single crystal at room
temperature and at 77 K. The room temperature trigonal symmetry was, as
expected, broken at low temperature.
\begin{figure}
  \centering
  \includegraphics[width=8.5cm]{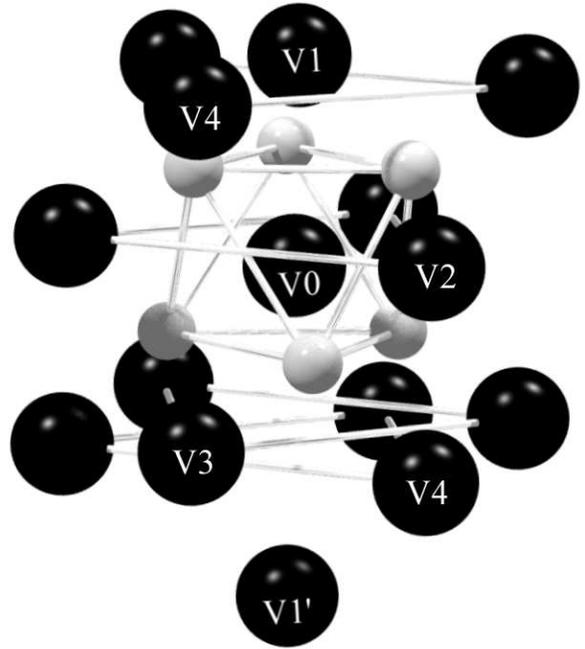}
  \caption{\label{structure} Local structure of V$_2$O$_3$. Black (grey)
           spheres represent vanadium (oxygen) ions. Only those oxygen
           ions which form an octahedron around the central vanadium atom
           V$_0$ are shown. The hexagonal $c$ axis is along the line
           V$_{1'}$-V$_0$-V$_1$.}
\end{figure}

The EXAFS measurements were performed at beam line X23B at the National
Synchrotron Light Source at Brookhaven National Laboratory. Spectra at
different temperature points above and below the transition temperatures of
the pure and doped sample were taken using a closed cycle helium cryostat. The
sample was held under vacuum to reduce thermal leakage and air-absorption and
prevent water condensation. The x-ray absorption spectra were measured in
fluorescence yield mode using a Lytle detector;\cite{Lytle84} background
radiation was filtered by a Ti foil. In fluorescence mode self-absorption
effects may strongly damp the EXAFS amplitude even at normal incidence. As the
primary effect of self-absorption is to reduce the EXAFS
amplitude,\cite{Troeger92, Pfalzer99} but leave the phase unchanged, the
distances of neighboring atoms obtained by EXAFS, which are determined largely
by the phase, are unaffected. We corrected for self-absorption using a
generalization of the method of Tr\"{o}ger \textit{et al.}\cite{Troeger92,
Goulon82}\ to large detector surfaces, as described elsewhere.\cite{Pfalzer99}
This correction is quite reliable, as was shown in Ref.~\onlinecite{Pfalzer99}
by demonstrating that for Al-doped V$_2$O$_3$ the absorption was isotropic in
the $ab$ plane after the correction, as expected for trigonal symmetry.

For the EXAFS measurements the sample was oriented so that the $(11\bar{2}0)$
plane (in hexagonal notation) was the surface perpendicular to the incident
beam. In this geometry the orientation of the polarization vector $\vec{E}$ of
the incoming x-rays with respect to the hexagonal $c$ axis can be changed by
rotating the sample around the surface normal. Measurements were made with
$\vec{E}$ parallel and perpendicular to $\vec{c}_{hex}$. As described by
Frenkel \textit{et al.},\cite{Frenkel97} this facilitates measuring the two
different components of the anisotropic absorption coefficient:
\[
\mu = \mu_\perp sin^2\theta + \mu_\parallel cos^2\theta ,
\]
where $\mu_\perp$ ($\mu_\parallel$) is the absorption coefficient for
$\vec{E}$ perpendicular (parallel) to $\vec{c}_{hex}$ and $\theta$ is the
angle between $\vec{E}$ and $\vec{c}_{hex}$. Since different scattering paths
contribute to $\mu_\perp$ and $\mu_\parallel$, measurement of these two
independent quantities achieves better separation of paths. After correcting
for self-absorption as described above standard EXAFS analysis was performed.
For the Fourier transform a window in $k$ space from 3 \AA$^{-1}$ to 12
\AA$^{-1}$ was used ($k$ being the wavevector of the photoelectron).

Measured spectra were compared to model spectra calculated with
FEFF8.\cite{Ankudinov98} The calculation requires estimated atomic positions,
provided by models of the structure. To construct these structural models we
used the lattice parameters provided by our XRD measurements. Then, for the
trigonal model, we used the atomic positions tabulated by
Wyckoff\cite{Wyckoff63} for the pure compound. For the monoclinic model, we
started with the measured hexagonal lattice parameters and tilted the
$c_{hex}$-axis by 1.995$^\circ$, the same amount as in the AFI phase of
undoped V$_2$O$_3$,\cite{Urbach95} to reproduce the monoclinic distortion. The
resulting pseudo-hexagonal lattice vectors were converted to monoclinic ones
using the conversion matrix of Ref.~\onlinecite{DM70}. Relative atomic
positions for the monoclinic phase were also taken from
Ref.~\onlinecite{DM70}. The structure refinement was performed by varying the
interatomic distances to all vanadium atoms up to the fourth nearest neighbor,
as shown in Fig.~\ref{structure}, as well as to the nearest two shells of
oxygen atoms. A few double scattering paths with high scattering amplitudes
were included in the fits. The nearest oxygen shell forms a (distorted)
octahedron around the absorbing vanadium atom V$_0$ and is responsible for the
well-separated peak of the Fourier transformed spectra between 1 and 2 \AA.
Fitting was carried out in $r$ space using $k^3$ weighting.
\begin{figure}
  \centering
  \includegraphics[width=8.5cm]{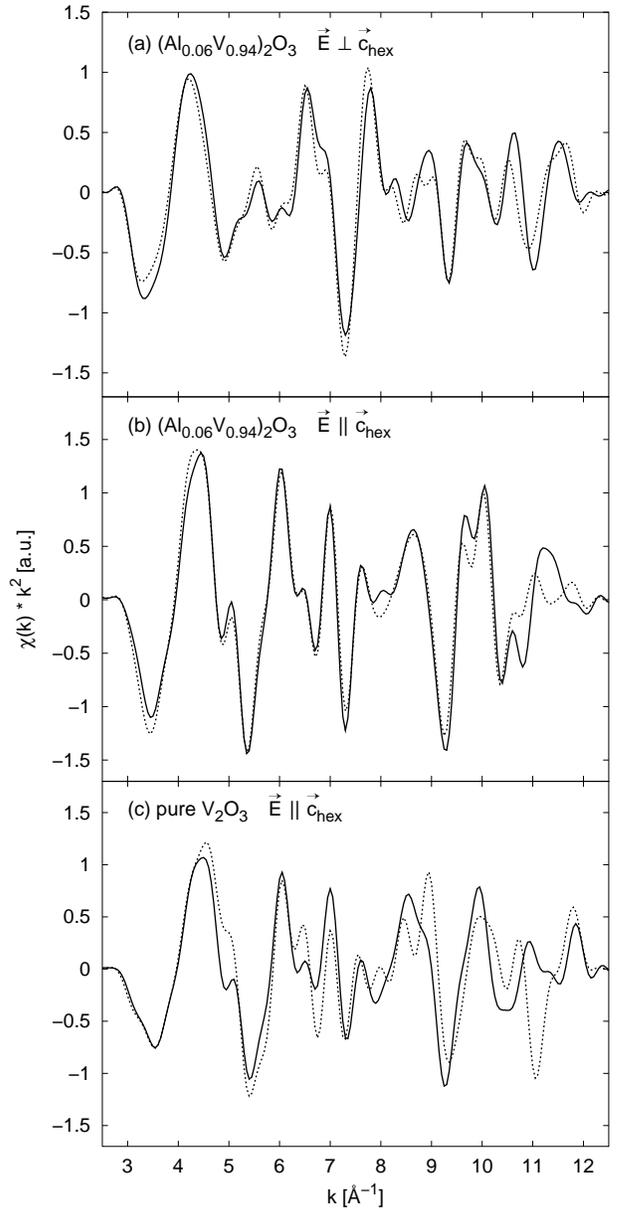}
  \caption{\label{EXAFS-ksp} EXAFS spectra of (Al$_{0.06}$V$_{0.94}$)$_2$O$_3$
           in $k$-space. The spectra above (dashed line) and below (solid) the
           PI to AFI transition are virtually the same for Al-doped V$_2$O$_3$
           for both $\vec{E} \perp \vec{c}_{hex}$ and $\vec{E} \parallel
           \vec{c}_{hex}$ (a and b), while pronounced changes are observed in
           pure V$_2$O$_3$ which is in the PM state at room temperature (c),
           showing the two insulating phases have virtually identical local
           physical structure which differs from that of the metal.}
\end{figure}

NEXAFS measurements on the O $1s$ edge were performed under ultra high vacuum at
the U41-1/PGM beamline at the BESSY 2 storage ring using the same crystal in
the same geometry as for the EXAFS measurements. The signal was monitored by
measuring the total electron yield; this is somewhat surface-sensitive, so
care was taken to prepare and maintain clean surfaces characteristic of the
bulk.

The experiments were complemented by electronic structure calculations based
on density functional theory (DFT) in the local density approximation (LDA).
The calculations used the augmented spherical wave (ASW) method in its
scalar-relativistic implementation,\cite{wkg,revasw} which was already
applied in Ref.~\onlinecite{Held01}. The present study included the paramagnetic
metallic V$_2$O$_3$, paramagnetic insulating
(V$_{0.962}$Cr$_{0.038}$)$_2$O$_3$, and antiferromagnetic insulating
V$_2$O$_3$ phase using the crystal structure data of Dernier
\cite{Dernier70,remark} as well as Dernier and Marezio.\cite{DM70}
Note that, since DFT is a ground state theory, differences between the phases
were taken into account only via the different crystal structure. In order to
study the effect of spin-polarization separately from the monoclinic
distortion in the AFI state, a complementary set of calculations with enforced
spin-degeneracy was performed for this structure.

In order to account for the openness of the crystal structures, empty
spheres, i.e.\ pseudo atoms without a nucleus, were included to model
the correct shape of the crystal potential in large voids. Optimal empty
sphere positions and radii of all spheres were automatically determined
by the recently developed sphere geometry (SGO) algorithm.\cite{vpop}
As a result, 8 and 16 empty spheres with radii ranging from 1.78 to 2.42
$a_B$ were included in the trigonal and monoclinic cell, respectively,
keeping the linear overlap of vanadium and oxygen spheres below 16.5\%.
The basis set comprised V $4s$, $4p$, $3d$ and O $2s$, $2p$ as
well as empty sphere states. Fast self-consistency was achieved by an
efficient algorithm for convergence acceleration.\cite{mixpap} Brillouin
zone sampling was done using an increasing number of $\vec{k}$ points
ranging from 28 to 2480 and 108 to 2048 points within the respective
irreducible wedges, ensuring convergence of the results with respect to
the fineness of the $k$-space grid.

\section*{Results}
In Fig.~\ref{EXAFS-ksp} we compare the EXAFS (in $k$ space) of a pure
V$_2$O$_3$ sample and the Al-doped sample above (dashed lines) and below
(solid lines) their respective transition temperatures. For the Al-doped sample
EXAFS spectra were measured with the polarization vector $\vec{E}$ oriented
along and perpendicular to the hexagonal $c$-axis. The EXAFS is very similar
for both orientations for both the PI and the AFI phase of this sample [see
Fig.~\ref{EXAFS-ksp}(a) and 2(b)], indicating that this transition, involving
a long range monoclinic distortion and magnetic order, is accompanied by only
minor changes in local structure. In contrast, for pure V$_2$O$_3$, the large
differences in the EXAFS oscillations apparent in Fig.~\ref{EXAFS-ksp}(c) show
that the local environment of the absorbing atoms is very different above and
below the MIT. For the PI and AFI phases of the Al-doped sample, small
differences are apparent at high $k$; these presumably are due to the distinctly
different Debye-Waller factors at the measuring temperatures of 30 K and 180
K, respectively.
\begin{figure}
  \centering
  \includegraphics[width=8.5cm]{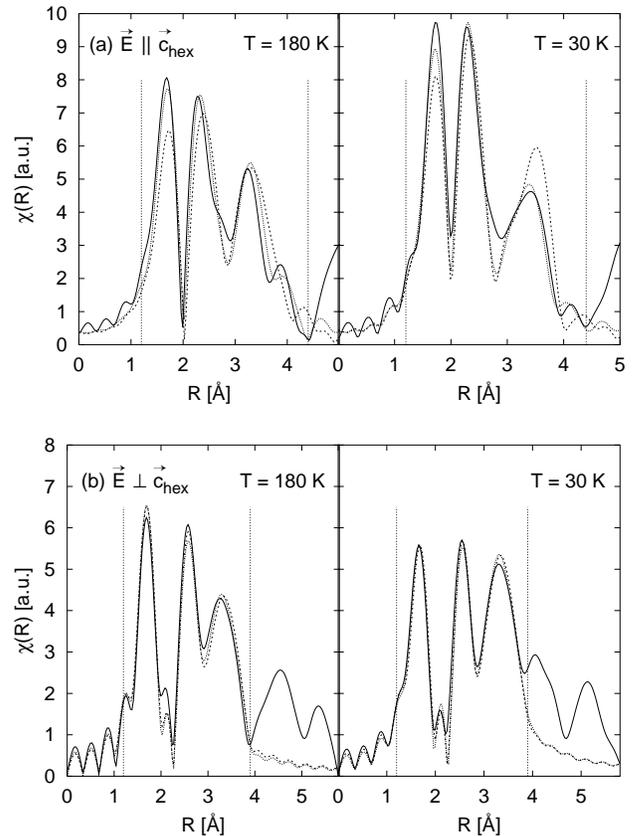}
  \caption{\label{EXAFS-rsp} FT Magnitudes of the EXAFS spectra of
           (Al$_{0.06}$V$_{0.94}$)$_2$O$_3$ (solid line) compared to
           calculated EXAFS using a monoclinic model (dotted lines) and a
           trigonal model (long dashes), for two geometries:
           (a) $\vec{E}$ parallel to $\vec{c}_{hex}$, in which significant
           differences between the models are expected. The monoclinic model
           fits the data much better than the trigonal one in both the PI
           (T = 180 K) and AFI (T = 30 K) phases. (b) $\vec{E}$ perpendicular
           to $\vec{c}_{hex}$. In this geometry many paths contribute and
           little difference is expected between the models, as observed.
           Dotted vertical lines indicate the R range used in the fit. The
           positions of the peaks are not corrected by the scattering phase
           shifts ($0.3-0.5$\ \AA ).}
\end{figure}

Fitting procedures, further discussed below, provide local interatomic
distances which confirm that the local structure is virtually the same in the
PI and AFI phase of the Al-doped sample, but also in the AFI phase of pure
V$_2$O$_3$. The fit also shows that both the PI and AFI phases of the Al-doped
V$_2$O$_3$ have a local symmetry which corresponds to a monoclinic lattice.
Fig.~\ref{EXAFS-rsp} compares the measured EXAFS (in $r$ space, where $r$ is
the interatomic distance) for both orientations of $\vec{E}$ to
$\vec{c}_{hex}$ to spectra derived from model calculations for monoclinic and
trigonal structures. It is evident that the monoclinic model fits the data
better. The trigonal model (long dashes) fits the data very poorly in the PI
phase at 180 K as well as in the AFI phase at 30 K, while the monoclinic model
(dotted lines) fits the data very well  in both cases, especially doing a much
better job in describing the distorted oxygen octahedron (peak at 1.7 \AA) and
the next vanadium neighbors (V$_1$ and V$_{1'}$ contributions are around 2.2
\AA\ and below 4 \AA\ in Fig.~\ref{EXAFS-rsp}). This visual observation is
confirmed by the fact that the reduced chi-squared of the trigonal model is
38.1, while that of the monoclinic model is 10.7. The better fit of the
monoclinic model than the trigonal model is particularly apparent in the
$\vec{E}$ parallel to $\vec{c}_{hex}$ orientation. This is expected because
the EXAFS measured in this orientation is more sensitive to the structural
differences between the phases; the major effect of the reduction in symmetry
during the transition from PI or PM to AFI is a tilt of the $c_{hex}$ axis and
a ``rotation'' of V$_0$-V$_1$ next neighbor pairs all located along this axis.
This orientation also achieves good separation of scattering paths, since
there is only a single V next neighbor atom (V$_1$ in Fig.~\ref{structure})
along the $c_{hex}$ axis. In fact, the main contribution to the scattered
intensity at 2.2 \AA\ in Fig.~\ref{EXAFS-rsp}(a) is due to the V$_0$-V$_1$
single scattering path. The broad peaks in the spectra taken with the
polarization vector $\vec{E}$ of the x-rays perpendicular to the $c_{hex}$
axis [Fig.~\ref{EXAFS-rsp}(b)] contain a large number of paths and therefore
fit equally well to both models.
\begin{table}
\begin{ruledtabular}
\begin{tabular}{c l c c c c c c}
  \multicolumn{2}{c}{ } &
  \multicolumn{4}{c}{(Al$_{0.06}$V$_{0.94}$)$_2$O$_3$} &
  {V$_2$O$_3$+Cr} & {V$_2$O$_3$} \\
  \multicolumn{2}{c}{Ion} &
  \multicolumn{2}{c}{ } & \multicolumn{2}{c}{measured} & &
  \\
  \multicolumn{2}{c}{(see Fig.~\ref{structure})} &
  \multicolumn{2}{c}{\raisebox{1.5ex}[-1.5ex]{calculated}} &
  \multicolumn{2}{c}{by EXAFS} & & \\
  \multicolumn{2}{c}{ } &
  trig. & mon. & PI & AFI & PI & AFI \\ \hline
  V$_1$ & & 2.72 & 2.72 & 2.76 & 2.79 & 2.75 & 2.74 \\
  \hline
   & i & & 2.86 & 2.96 & 2.91 & & 2.86 \\
  V$_2$ & ii & 2.91 & 2.87 & 2.97 & 2.93 & 2.92 & 2.88 \\
   & iii & & 3.00 & 3.10 & 3.05 & & 2.99 \\ \hline
   & i & & 3.43 & 3.44 & 3.42 & & 3.44 \\
  \raisebox{1.5ex}[-1.5ex]{V$_3$} & ii & \raisebox{1.5ex}[-1.5ex]{3.44} &
    3.45 & 3.46 & 3.44 & \raisebox{1.5ex}[-1.5ex]{3.45} & 3.46 \\ \hline
   & i & & 3.62 & 3.69 & 3.72 & & 3.63 \\
  V$_4$ & ii & 3.69 & 3.71 & 3.78 & 3.81 & 3.70 & 3.73 \\
   & iii & & 3.72 & 3.79 & 3.82 & & 3.74 \\
\end{tabular}
\end{ruledtabular}

  \caption{\label{distances}Distances in \AA\ from the central ion V$_0$ to nearby
           V ions (as labeled in Fig.~\ref{structure}) for various V compounds.
           The calculated values for (Al$_{0.06}$V$_{0.94}$)$_2$O$_3$ are obtained
           using hexagonal lattice vectors measured by XRD and relative atomic
           positions, assuming a trigonal (trig)\cite{Dernier70} and monoclinic
           (mon)\cite{DM70} lattice. (For the monoclinic lattice the ions V$_2$,
           V$_3$ and V$_4$ become non-degenerate and hence are identified
           separately in the table.) The values measured by EXAFS for the PI
           and AFI phases are those obtained using the fits shown in
           Fig.~\ref{EXAFS-rsp}. Distances for V$_2$O$_3$+Cr, specifically
           (Cr$_{0.038}$V$_{0.962}$)$_2$O$_3$, in the PI phase were measured by
           Dernier.\cite{Dernier70} Values for the AFI phase of pure V$_2$O$_3$ are
           calculated from the monoclinic lattice vectors and atomic positions
           published by Dernier and Marezio.\cite{DM70}}
\end{table}

Table~\ref{distances} lists the distances from a central reference ion,
labeled V$_0$ in Fig.~\ref{structure}, to nearby V ions for the various phases
of the V$_2$O$_3$ compounds. The values measured by EXAFS for the PI and AFI
phases are those obtained using the fits shown in Fig.~\ref{EXAFS-rsp}. For
comparison, we include interatomic distances calculated for
(Al$_{0.06}$,V$_{0.94}$)$_2$O$_3$ obtained using hexagonal lattice vectors
measured by XRD and relative atomic positions, assuming a
trigonal\cite{Dernier70} and monoclinic lattice.\cite{DM70} Distances for
(Cr$_{0.038}$V$_{0.962}$)$_2$O$_3$ in the PI phase were measured by
Dernier\cite{Dernier70} using XRD and those for the AFI phase of pure
V$_2$O$_3$ are calculated from the monoclinic lattice vectors and atomic
positions published by Dernier and Marezio.\cite{DM70} Uncertainties are
typically $\pm$0.02 \AA. It is evident that the interatomic distances measured
by EXAFS for the PI and AFI phases (Al$_{0.06}$V$_{0.94}$)$_2$O$_3$ are
essentially the same. In addition, these distances are in correspondence with,
though consistently slightly larger than, those obtained for the AFI phase of
pure V$_2$O$_3$, confirming that all insulating phases have essentially the
same local structure.
\begin{figure}
  \centering
  \includegraphics[width=8.5cm]{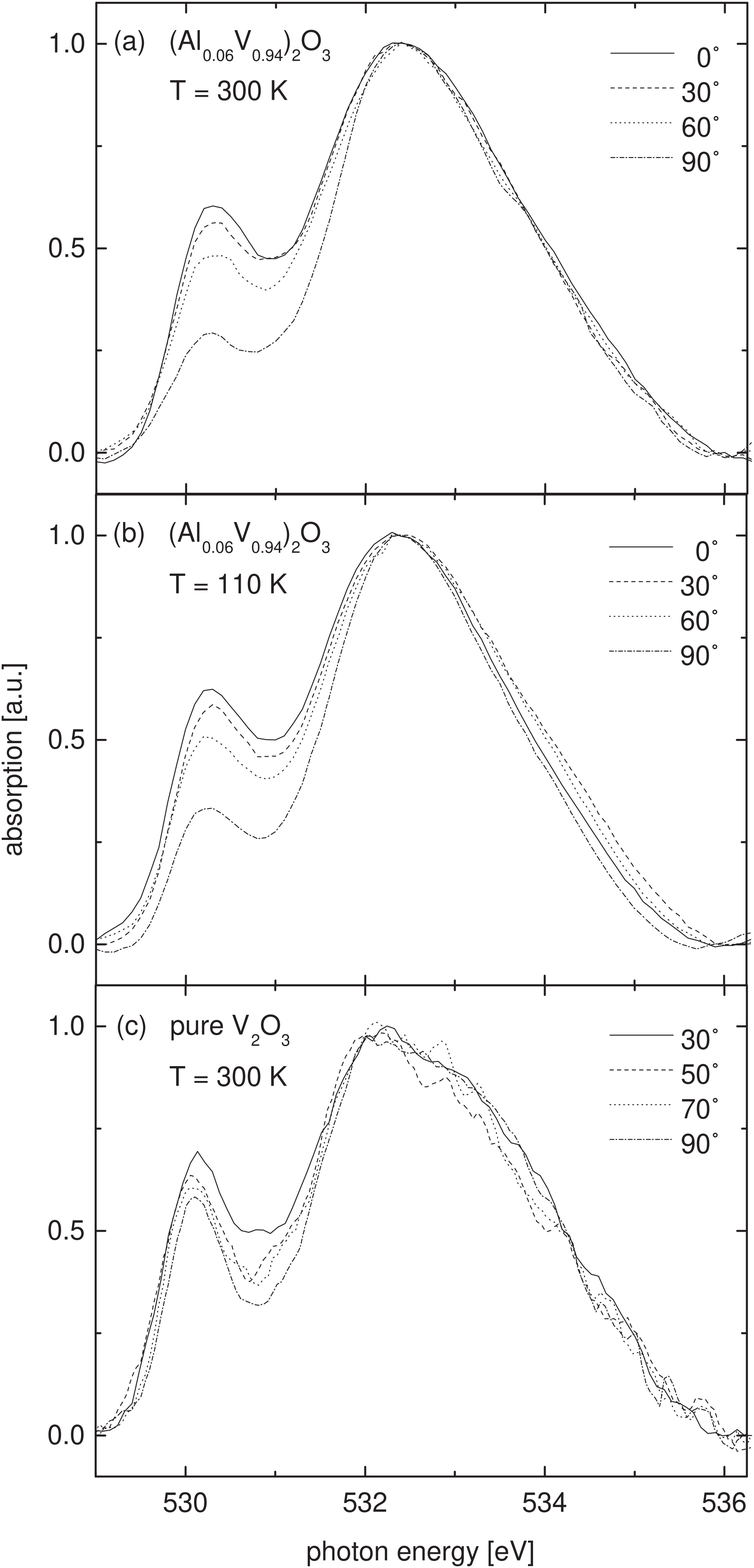}
  \caption{\label{NEXAFS-O1s} NEXAFS spectra of the O $1s$ edge of
           (Al$_{0.06}$V$_{0.94}$)$_2$O$_3$ as a function of angle between
           $\vec{E}$ and the $c_{hex}$ axis, showing large but identical
           angular anisotropy for (a) the PI and (b) the AFI phase. For
           comparison, this angular dependence almost vanishes for pure
           V$_2$O$_3$ in its metallic state (c) (from Ref.~\onlinecite{Mueller97},
           energy-shifted to facilitate comparison).
           All spectra were normalized to the large maximum at about 532 eV.}
\end{figure}

NEXAFS spectra of the O $1s$ edge in the same Al-doped Al$_2$O$_3$ crystal are
presented in Fig.~\ref{NEXAFS-O1s} (a) and (b). Large changes in the spectra are
apparent as the sample is rotated from a geometry with $\vec{E}$ parallel to
$\vec{c}_{hex}$ ($\vartheta=0^\circ$) to one with $\vec{E}$ perpendicular to
$\vec{c}_{hex}$ ($\vartheta=90^\circ$). It is evident that cooling from the PI
to the AFI phase has no significant effect on the spectra or their dependence
on angle. These spectra and their angular dependence are similar to those of
pure V$_2$O$_3$ in its low temperature AFI phase,\cite{Mueller97} but
different from those of pure V$_2$O$_3$ in its high temperature PM phase,
which has a much weaker angular dependence [Fig.~\ref{NEXAFS-O1s}(c)].

\section*{Discussion}
The EXAFS and O edge soft x-ray absorption measurements presented here
consistently show that the structure, at least on a local scale, and the V-O
hybridization in the PI and AFI phase of (Al,V)$_2$O$_3$ are similar, but
distinctly different from the metallic phase. The most obvious difference
between the metallic and insulating phases is the anisotropy of the O $1s$
x-ray absorption spectra, reflecting the V $3d$ - O $2p$ hybridization. These
findings suggest that the structural change at the MI transition (long or
short range order) is directly connected with the emergence of the insulating
state, e.g.\ via changes of hybridization.
\begin{figure}
  \centering
  \includegraphics[width=8.5cm]{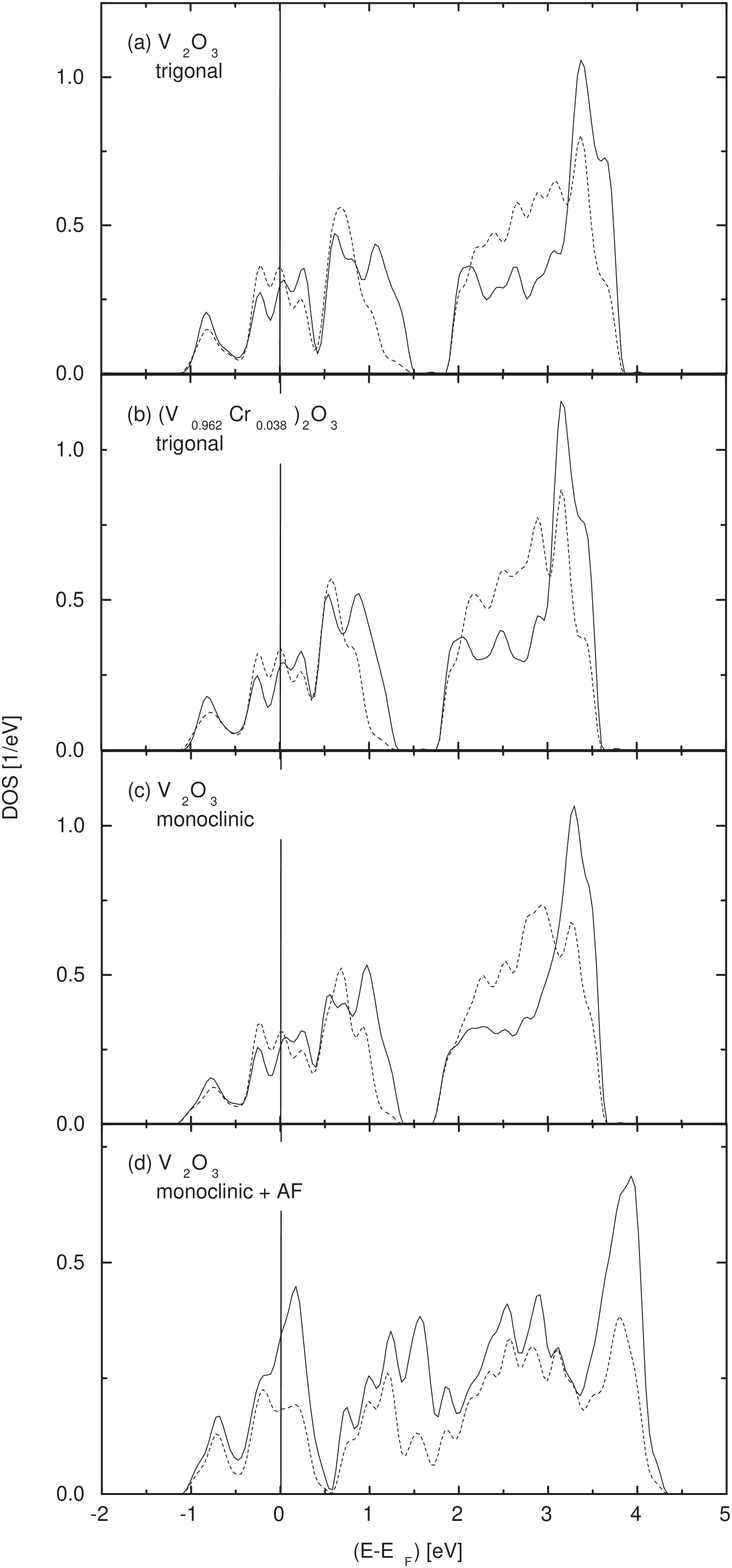}
  \caption{\label{LDA} Partial oxygen $2p$ densities of states predicted by
           the density functional calculations described in the text for
           different crystal structures of (V$_{1-x}$Cr$_x$)$_2$O$_3$ with
           various c/a ratios and symmetries: (a) the trigonal paramagnetic
           phase ($x=0$); (b) paramagnetic insulating phase with a larger
           trigonal distortion corresponding to $x=0.038$; (c) the monoclinic
           paramagnetic phase of V$_2$O$_3$; and (d) the monoclinic
           antiferromagnetic phase. Solid (dashed) lines indicate O $2p_z$
           ($2p_x$ or $2p_y$) states. For the magnetically ordered phase (d)
           spin up and spin down states have been added to facilitate
           comparison and the vertical scale has been expanded.}
\end{figure}

To estimate the effect of purely structural changes on the hybridization, we
show in Fig.~\ref{LDA} partial (projected) O $2p_z$ and $2p_{x,y}$ densities
of states (DOS) predicted by the electronic structure calculations. The V $3d$
partial DOS predictions for the spin-degenerate cases (see
Ref.~\onlinecite{Held01}) are in good agreement with those of
Mattheiss\cite{mattheiss94} and Ezhov {\em et al.}\cite{Ezhov99} The V $3d$
and O $2p$ partial DOS presented in Fig.~\ref{LDA} are very similar for all
phases, except for a slight decrease of the O $2p$ band width in the PI phase
as compared to the PM phase. In particular, the experimentally observed
optical band gap is not reproduced by the calculations and all phases are
predicted to be metallic. These discrepancies between experiment and
calculations are usually attributed to the fact that on-site electron-electron
correlations are not fully taken into account by LDA. The calculations also
predict that the O $2p$ DOS of all phases is isotropic, i.e.\ the projected
$2p_z$ and $2p_{x,y}$ DOS are similar, implying that NEXAFS spectra of the O
$1s$ edge with the polarization vector $\vec{E}$ parallel or perpendicular to
$\vec{c}_{hex}$ should also be alike. Experimentally, this is only observed
for the metallic phase of V$_2$O$_3$, as shown in Fig.~\ref{NEXAFS-O1s}(c),
but not for the insulating phases. However, the calculations for
(V$_{0.962}$Cr$_{0.038}$)$_2$O$_3$ [Fig.~\ref{LDA}(b)] described above did not
include the local distortions observed by EXAFS in the PI phase and the
calculations for monoclinic V$_2$O$_3$ [Fig.~\ref{LDA}(c)] artificially
enforced a spin-degenerate state. If the antiferromagnetic ground state of the
monoclinic phase is taken into account by using a spin-polarized calculation
[Fig.~\ref{LDA}(d)], the partial DOS displays various band shifts, although
the experimentally observed optical band gap is still not reproduced. In
particular, the O $2p_z$ partial DOS now differs substantially from the
$2p_{x,y}$  near 0.3 eV, 1.7 eV, and 4 eV and is, therefore, more compatible
with the observed  anisotropy of the O $1s$ spectra. In this context it should
be noted that according to neutron scattering measurements antiferromagnetic
short range order persists in both the PI and the PM phase, although the
anisotropy of the O $1s$ spectra is only observed in the PI and AFI phases.

To address the effect of electronic correlations on the electronic structure
of V$_2$O$_3$, recently LDA was combined with dynamical mean field theory
(DMFT) and calculated and measured photoemission and x-ray absorption spectra
were compared on the basis of the calculated V $3d$ spectral
weight.\cite{Held01} The electron-electron interaction shifts the $a_{1g}$ and
$e_g^\pi$ states with respect to each other. This would cause an anisotropy in
the oxygen $1s$ spectra, since the $a_{1g}$ and $e_g^\pi$ states hybridize
differently with the O $2p_z$ and $2p_{x,y}$ states. According to LDA the V
$a_{1g}$ hybridize primarily with the O $2p_z$ orbital along the $c_{hex}$
axis, while the $e_g^\pi$ hybridize primarily with the O $2p_{x,y}$ in–plane
orbitals. Therefore polarization-dependent excitations of O $1s$ core
electrons into unoccupied O $2p_z$ and O $2p_{x,y}$ states provide information
on their hybridization with V $a_{1g}$ and $e_g^\pi$ states respectively.
Based on this we concluded earlier\cite{Mueller97}that the observed anisotropy
in the AFI phase of pure V$_2$O$_3$ results from an increase in $a_{1g}$
character in the unoccupied DOS, causing a corresponding increase of  weight
of O $p_z$ hybridized states above  $\epsilon_F$. This interpretation is
consistent with the conclusion of Park,\cite{Park00} based on measurements of
the V $L_{2,3}$ edges, that there is an increase in $e_g^\pi$ occupancy during
the transition from the metallic to the insulating state in V$_2$O$_3$. On the
other hand, the relative shifts of the $a_{1g}$ and $e_g^\pi$ states predicted
from LDA + DMFT calculations are not sufficient to account for the observed
anisotropy of the O $1s$ spectra in the insulating state.

Recently Shiina\cite{Shiina01} attributed the monoclinic lattice distortion in
the AFI phase to orbital ordering which would make the three originally
equivalent magnetic bonds in the $ab$-plane inequivalent and cause a
monoclinic lattice distortion. Evidence for orbital fluctuations in the PI
phase was provided by neutron scattering measurements,\cite{Bao98} which
showed that magnetic short range order was limited to nearest neighbor
distances, resulting in a first order transition from the PI to the AFI phase.
Given the results of Ref.~\onlinecite{Shiina01}, orbital fluctuations in the
PI phase could cause a dynamic monoclinic distortion. Assuming the time scale
for such fluctuations is long compared to the time scale of the x-ray
absorption process, EXAFS and soft x-ray absorption would measure an
instantaneous structure and the PI phase would appear monoclinic, while XRD,
which measures on a much longer time scale, would see a trigonal lattice. In
the AFI phase the monoclinic distortion might become static although, on a
local scale, still the same as in the PI phase. This model would account for
the fact that neither EXAFS nor soft x-ray absorption observe differences
between the PI and the AFI phase, in contrast to XRD. The spectroscopic
results are, however, also consistent with static short range orbital order in
the PI phase, which would, accordingly, be an orbital glass, with an
disorder-order transition to the ordered AFI phase.
\newline
The local monoclinic distortion we have found in
(V$_{0.94}$Al$_0.06$)$_2$O$_3$ is reminiscent of the much smaller, but
significant monoclinic distortion found in the metallic phase of pure
V$_2$O$_3$.\cite{Frenkel97} From the fact that this monoclinic distortion is
not detected in XRD measurements, Frenkel \textit{et al.}\cite{Frenkel97}\ set
an upper limit of 40 \AA\ on the size of possible monoclinic domains and
concluded that the MIT contains both an order-disorder and a displacive
component. The monoclinic distortion in the metallic phase was determined to
be about 30\% of that in the antiferromagnetically ordered insulating phase.
From the above discussion it can be concluded that their data suggest orbital
fluctuations are also present in the metallic phase of V$_2$O$_3$, although
local distortions are less prominent than in the PI phase, possibly due to
better screening in the metallic phase.

\section*{Conclusion}
The x-ray absorption measurements presented here show that both the PI and AFI
insulating phases of V$_2$O$_3$ are distinguished from the PM phase by: (i)
the presence of local or long range distortion of the lattice (probably
connected to short or long range orbital order, respectively) and (ii)
differences in the V $3d$ - O $2p$ hybridization, accompanied by corresponding
band shifts. Both the distortion and the hybridization appear to be
independent of the presence of antiferromagnetic correlations, which are
present in all phases.\cite{Bao98} The similarity of the PI and the AFI
phases, at least on a local scale, suggests a common route from their
insulating behavior to the metallic behavior of the PM phase. Interactions
between orbital degrees of freedom, which lead to an orbitally ordered state
in the AFI phase and orbital short range order in the PI phase, appear to be
an an important fingerprint of the MIT. The characteristic differences between
V $3d$ - O $2p$ hybridization in the metallic and the insulating phases
suggest that those changes in hybridization play a role in the MIT. Such
changes might be due to strong anharmonic contributions to the
temperature-dependent phonon spectrum.

\begin{acknowledgments}
We appreciate valuable assistance in the measurements and analysis from J.\
Kirkland at NSLS and Ch.\ Jung and M.\ Mast at BESSY and enlightening
discussions with P.\ Riseborough and K.-H.\ H\"{o}ck. This work was supported in
part by the BMBF under contract number 0560GWAA and the DFG under contract
number HO-955/2 and SFB484 and by the US DOE Contract DEFG02-91-ER45439. The
NSLS is supported by the DOE.
\end{acknowledgments}

\bibliography{AlV2O3}

\begin{thebibliography}{30}
\bibitem[{rem()}]{remark}
\bibinfo{note}{{U}se of the crystal structure of Cr-doped V$_2$O$_3$ for the
  paramagnetic insulating phase is justified by the observation that Cr doping
  has very similar effects as Al doping and is equivalent to the application of
  (negative) pressure.}
\end{thebibliography}
\end{document}